# Characterisation of local ICRF heat loads on the JET ILW

P. Jacquet[a]*, F. Marcotte[b], L. Colas[c], G. Arnoux[a], V. Bobkov[d], Y. Corre[c], S. Devaux[d], J-L Gardarein[e], E. Gauthier[c], M. Graham[a], E. Lerche[f], M-L. Mayoral[a], I. Monakhov[a], F. Rimini[a], A. Sirinelli[a], D. Van Eester[f], and JET EFDA contributors[†]

*JET-EFDA, Culham Science Centre, Abingdon, OX14 3DB, UK.*
[a]*Euratom/CCFE Fusion Association, Culham Science Centre, Abingdon, OX14 3DB, UK.*
[b]*Ecole Nationale des Ponts et Chaussées, F77455 Marne-la-Vallée, France.*
[c]*CEA, IRFM, F-13108 Saint-Paul-Lez-Durance, France.*
[d]*Max-Planck-Institut für Plasmaphysik, EURATOM-Assoziation, Garching, Germany.*
[e]*Aix-Marseille University, IUSTI UMR 6995, 13013 Marseille, France.*
[f]*Association EURATOM-Belgian State, ERM-KMS, Brussels, Belgium.*

**Abstract**

When using Ion Cyclotron Range of Frequency (ICRF) heating, enhanced heat-fluxes are commonly observed on some plasma facing components close to the antennas. Experiments have recently been carried out on JET with the new ITER-Like-Wall (ILW) to characterize the heat flux to the JET ICRF antennas. Using Infra-Red thermography and thermal models of the tiles, heat-fluxes were evaluated from the surface temperature increase during the RF phase of L-mode plasmas. The maximum observed heat-flux intensity was ~ 4.5 MW/m$^2$ when operating with -π/2 current drive strap phasing at power level of 2MW per antenna and with a 4 cm distance between the plasma and the outer limiters. Heat-fluxes are reduced when using dipole strap phasing. The fraction of ICRF power deposited on the antenna limiters or septa was in the range 2-10% for dipole phasing and 10-20% with +/-π/2 phasing.

PACS: 52.55.Fa, 52.50.Qt, 52.35.Mw, 52.40.Fd, 52.40.Kh

*Corresponding author address: Euratom/CCFE Fusion Association, J20/1/08, Culham

---

[†] See the Appendix of F. Romanelli et al., Proceedings of the 23$^{rd}$ IAEA Fusion Energy Conference 2010, Daejeon, Korea.

Science Centre, Abingdon, OX14 3DB, UK

*Corresponding Email: philippe.jacquet@ccfe.ac.uk

Presenting author: Philippe Jacquet

Presenting author Email: philippe.jacquet@ccfe.ac.uk

1 Introduction

During the application of Ion Cyclotron Range of Frequency (ICRF) heating, local heat deposition in the vicinity of the antennas is commonly observed as was reported in Tore-Supra [1] and JET [2]. This phenomenon is attributed to the interaction of ICRF waves with the SOL plasma through RF sheath rectification [3][4]. Infra-red (IR) thermography was used with the JET carbon (C) wall to quantify heat-fluxes on the protection septa of the JET A2 antennas [2]. Surface temperature increases (~ 600 $^{o}$C in some cases) were explained by large temperature gradients in layers with poor thermal contact with the bulk tiles, while these Plasma Facing Components (PFCs) were submitted to modest heat-fluxes (few MW/m$^2$). Since 2011, JET has operated with a new ITER-like wall (ILW) consisting mainly of beryllium (Be) tiles in the chamber and tungsten tiles in the divertor [5]. Experiments were carried out recently to further characterize local ICRF heat deposition in JET with the following goals:

(a) Quantify the heat-fluxes on the tiles surrounding the antennas using IR thermography.

(b) Verify that the heat fluxes associated with ICRF is within the PFC thermal handling capabilities, in particular to avoid melting.

(c) Characterise spatial pattern and intensity parametric dependence of hot-spots in order to better understand RF sheath rectification, the long term goal being the optimisation of ICRF antenna designs [6] to minimize the intensity of $E_{//}$ the RF electric field component parallel to the (static) magnetic field at the antenna aperture. This $E_{//}$ field drives rectification of the sheath potential on PFCs magnetically connected to the antenna [7], which has detrimental consequences on local power dissipation and on impurity generation.

(d)   Provide experimental data that could be used for extrapolation to the ITER ICRF system.

This paper focuses on the above points (a) and (b); points (c) and (d) are still subject of active studies, their progress will be reported in future publications.

## 2   JET A2 antenna system and wide angle IR camera

Figure 1 shows a schematic top view of the JET tokamak. The A2 ICRF antennas A, B, C, and D and the ITER-Like ICRF antenna (ILA, not used in 2011) are represented as well as the camera view which covers antenna A, the ILA antenna and half of antenna B. Each A2 antenna [8][9] is a phased array of 4 poloidal straps; controlling the phase between straps allows waves to be launched with different $k_{//}$ spectra. Usually π (dipole phasing) or +/- π/2 phasing (current drive phasings) between adjacent straps are used. The plasma facing part of the antennas is covered by a Faraday screen consisting of tilted solid Be rods. Each antenna is surrounded by two poloidal limiters made of solid Be tiles [10], and a vertical septum made of solid Be is fitted at the centre (between straps 2 and 3) of each antenna. In the experiments described here straps 1&2 of A2 antennas A and B were fed by the same RF amplifiers through a 3dB hybrid coupler system [11]. Straps 3&4 of antenna A were fed by independent amplifiers. In 2011/12, straps 3&4 of antenna B were not used; the transmission lines were short-circuited at the generator end. The matching elements (stubs and trombones) were detuned to make sure that at the operating frequency, the transmission lines feeding these straps were not acting as a high Q resonators.

The JET wide angle IR camera is described in [12]. The IR camera wavelength measurement range is 3.97-4.01 μm. The pixel size of the camera is approximately 1.5 cm at distances corresponding to the antenna A septum. For the pulses described here the camera time resolution was 20 ms (50 Hz acquisition rate) and the exposure time was 600 μs. A linear two point Non-Uniformity Correction (NUC) was applied to take into account the different responses (offset and gain) of each pixel [13]. In order to measure the Be wall temperature, the camera was calibrated as follows:

- Thermocouples (Tc) imbedded in some of the inner or outer wall Be tiles viewed by the camera were taken as temperature references.
- The temperature of the wall was increased step by step in a series of inner/outer limiter plasma pulses (RF not used in the outer limiter plamas).
- The IR camera was calibrated against the Tc measurements. Calibration data were recorded at the beginning of each JET pulse, before any heating was applied to the wall in order to ensure the thermal equilibrium of the Be tiles instrumented with Tcs (time between pulses at least 20 minutes).

The Be surface temperature is deduced from the IR measurements assuming that *DL* the response of each pixel is linear function of surface spectral emission in the camera wavelength range of measurement:

$$DL = a + \frac{b}{e^{hc/(\lambda kT)} - 1} \qquad (1)$$

where h is the Planck constant, c is the speed of light in vacuum, k is the Boltzmann constant, $\lambda$ is the centre of the camera wavelength range and T is the temperature in Kelvin. The calibration curve used to deduce the Be surface temperature from the IR signal is shown in Figure 2; the parameters *a* and *b* (Equation 1) were obtained from a fit procedure to the calibration data and extrapolation was used for the highest temperatures. This calibration is coherent with the Be emissivity in this wavelength range ($\varepsilon_{Be}$ ~0.18) as evaluated independently in [14]. The error bars on Figure 2 represent the uncertainties in surface temperature from the IR data; they are derived from the scatter in the calibration data. The systematic difference between the inner and outer poloidal tiles used for the calibration is likely due to slight differences in the surface emissivity of each tile.

## 3   Evaluation of heat-fluxes from surface temperature measurements

A linear deconvolution procedure [15] was used to determine the heat-flux on the Be PFCs seen by the camera from the time response of surface temperature. In this procedure, the tile thermal properties are taken into account through the surface temperature response to a Heaviside step excitation with a reference heat-flux. This is obtained from modelling. The tiles of the septa and of the poloidal limiters around the ICRF antennas were modelled using the ANSYS™,[‡] finite element software and with a simplified 1D finite difference model [16]. An example for a septum tile is shown in Figure 3. The thermal response can slightly vary depending on the size of the heated area, which was adjusted in the ANSYS simulations to match the size of the hot-spots on the IR camera view. The simplified 1D model and ANSYS thermal responses are close for analysis times of ~ 20 s or shorter; this correspondence is even better when modelling the outer poloidal limiters tiles because castellations at the surface of the tiles [10] (~12×13mm) prevent lateral heat diffusion. The 1D reference responses were used in the deconvolution procedure to calculate the heat-fluxes presented in this paper. Overall the error on heat-fluxes from the uncertainty on the tile thermal response is estimated to be +/- 5%.

Figure 4 shows an example of the heat-flux estimated on the antenna A septum when monitoring the temperature of the most intense hot-spot for pulse 81719, one of the pulses from the series described in section 4. In this pulse ($B_T$=2.5 T, Ip=2 MA, 3cm plasma outer-limiter distance in the equatorial plane), the ICRF power was launched by antenna A and B (2 MW launched ICRF power in total); only straps 1&2 were used with a strap phasing of $\Phi_2$- $\Phi_1 = -\pi/2$. In this case the hot-spot on the antenna A septum corresponds to a heat-flux of ~1.8 MW/m$^2$. The calculated heat-flux when the ICRF power was turned off (t>11s) is close to zero (and, in particular, it does not become negative) which is an indication of the quality of our thermal model. In particular there is no need to consider a surface layer poorly thermally attached to the tile, in contrary to the carbon wall analysis. The experiments were performed few months after JET operation with the ILW started, and no beryllium evaporation nor boronization were done for wall conditioning. The heat-flux 'in-rush' observed during the 1$^{st}$ second of the ICRF pulse corresponds to a phase where the antenna matching elements were moving and the strap feeding voltages (Figure 4-b) and phases (Figure 4-c) were not yet stabilized. In particular, there was a strong voltage in-balance in the transmission lines during this phase. This illustrates the importance of antenna strap feeding conditions in the RF sheath rectification and local heat deposition phenomena [6]. More studies are ongoing to understand the relation between strap feeding conditions and hot-spot intensities on JET, in particular by computing the $E_{//}$ RF field map at the antenna aperture using the TOPICA antenna modelling code [17].

---

[‡] ANSYS™, ANSYS Inc., Canonsburg, PA 15317, USA.

## 4    Characterisation of local ICRF heat-loads

An IR frame for pulse 81719 is shown in Figure 5. Hot-spots developed on the energized antenna septa and on the neighbouring limiters. These hot-spots were not observed when using remote antennas with the same amount of RF power and so cannot be attributed to fast ions escaping from the plasma. The exact hot-spot pattern can change, depending on: the antenna feeding conditions (strap phasing, number of straps energized); the plasma configuration, in particular on the antenna–plasma distance (which affects antenna loading and thus RF electric field at the antenna aperture); and the magnetic configuration which affects the field line incidence on the PFCs. A series of pulses were used to quantify the RF hot-spot intensity as a function of ICRF system and plasma parameters. In this series, straps 1&2 or 1,2,3&4 of antenna A were used together with straps 1&2 of antenna B. The ICRF power launched from antenna A and B was varied in the range 0.5-3 MW (27 kV max voltage in the transmission lines) with dipole or +/-π/2 strap phasing. The plasma parameters were: $B_T$ in the range 2.4-2.5 T; Ip=2 MA; L-mode plasma; 1$^{st}$ harmonic (H)D minority ICRF heating (~2% H concentration); and plasma outer-limiter distance in the range 3-6 cm. The ICRF frequency was 42 MHz and the H minority resonance was located between R=2.65m and R=2.75 m (on the high field side of the plasma centre located at R=2.96 m). The electron density in the SOL was measured with a reflectometer [18] (Figure 1 shows the reflectometer location: measurements are taken ~75º away from antenna A). We define as $n_{e,lim}$, the electron density at a mid-plane radius of 3.87 m (1 cm in front of the main outer poloidal limiters) During the experiment, $n_{e,lim}$ was changed (in the range $1.5 \times 10^{18} - 4.5 \times 10^{18}$ m$^{-3}$) by varying the outer radius of the plasma. Figure 6 summarizes the main dependencies of the ICRF hot-spots intensity. This Figure enables a simple assessment of the maximum hot-spot intensity, as a function of two controllable operating parameters which is useful for the purpose of the Be wall protection. This figure also illustrates some important aspects of the mechanisms leading to the formation of hot-spots although we do not intend to present here a scaling law describing all the parameters influencing RF sheaths rectification and local hot-spot formation. The most intense heat-flux around antenna A (the location could change from pulse to pulse depending on plasma and ICRF system configurations) is plotted as a function of $n_{e,lim} \times V^2$, $V$ is the ICRF voltage at the antinodes of the transmission lines feeding the straps, averaged over the active straps of antenna A *($V=<V^{foward}+V^{reflected}>$)*. The hot-spot intensity increases roughly as the electron density at the outer limiter position and as the square of the RF voltage in the transmission lines feeding the antenna. This behaviour is different from the one proposed in previous experiments with the JET carbon wall ($n_{e,lim} \times V$ as described in [2][19]), which would be coherent with simple RF sheath rectification models [20] where the intensity of the heat-flux on PFCs from RF sheath rectification is described by:

$$Q^{RF} \sim ne \cdot \sqrt{k(T_e + T_i)/m_i} \cdot |eV_{DC}| \qquad (2)$$

with k: Boltzmann constant; $T_e$ and $T_i$: respectively electron and ion temperature in front of the PCF; $m_i$: ion mass; $V_{DC}$: Rectified sheath potential. Assuming that $|V_{DC}|$ is proportional to the RF $E_{//}$ field integrated along the field line intercepted by the PFCs [7], and that $E_{//}$ is also proportional to the RF voltage on the antenna straps one could expect from simple RF sheath rectification models:

$$Q^{RF} \propto ne \cdot \sqrt{k(T_e + T_i)/m_i} \cdot \left| e \int_{fieldlines} E_{//} \right| \qquad (3)$$

$$Q^{RF} \propto ne \cdot \sqrt{k(T_e + T_i)/m_i} \cdot V \qquad (4)$$

The hot-spots intensity might be strongly influenced by local modifications of the plasma characteristics from the interaction with the ICRF field, for example by the RF driven E×B flows in front of the antenna [21] (it is reminded that the density measurements in the SOL are not taken directly in front the antennas). Also, local modifications of $T_e$ and $T_i$ when applying the ICRF power cannot be ruled out.

In some cases (black solid triangles) the hot spot intensity even exhibits a dependency with ~ $V^3$, the maximum hot-spot intensity (~4.5MW/m$^2$) being observed in this series of pulses. In this case all four straps of antenna A were used and the power launched by the antenna was ~2 MW. For these pulses it was observed that the antenna coupling resistance increased as the launched power was increased, pointing towards a net and localised (not seen from reflectometry) increase of the electron density in front of the antenna.

The hot-spot intensity increases when using more active straps on the antenna (or in other words when more power is launched for a given V). Increasing the number of active straps could change the $E_{//}$ field structure and in turns could increase $\int_{fieldlines} E_{//}$, or the net increase of launched power could also lead to stronger plasma properties modifications in front of the antenna.

The hot-spot intensity increases when using +/-π/2 strap phasing instead of dipole; asymmetric antenna phasings are indeed expected to increase $E_{//}$ [6].

For reference, C-wall measurements on septum A have been added in Figure 6 ($n_{e.,lim}$ in the range $1.3\times10^{18} - 2.6\times10^{18}$ m$^{-3}$, $V$ up to 20 kV for the carbon-wall data). The measurements with the Be wall confirm the order of magnitude of the heat-flux intensity (few MW/m$^2$) derived from the C-tiles thermal analysis in which we needed to include a model of poorly attached surface layers to explain large and fast temperature increases when applying ICRF [2]. The errors in the absolute values of heat flux can contribute importantly the differences in Be wall / C wall hot spot intensity visible in Figure 6. The error-bar for one of the carbon wall data point includes the uncertainties related to the thermal model, but not the uncertainty in the surface temperature measurement [2]. The error bar for the Be-wall data includes the uncertainties from both the temperature measurement and the thermal model. The different operating conditions between the Be-wall and the carbon-wall experiments (change in the layout of the transmission lines feeding straps 3&4 of antenna A, and also more generally change of SOL plasma properties) might also contribute to the observed difference in heat-flux intensity.

The fraction of ICRF power dissipated on neighbouring limiters for antennas A and B and the septa located in the IR camera view was also evaluated. During ICRF pulses, the heat-flux was evaluated on a regular grid covering these PFCs. The power dissipated on these objects, $P_{dissip}$, was then obtained by integrating the heat-flux over the surfaces. The fraction of ICRF power directly dissipated locally was calculated as:

$$\eta_{ICRF} = (P_{dissip} - P_{cond})/P_{ICRF} \qquad (5)$$

$P_{cond}$, the conducted plasma power falling on the same limiters was evaluated as:

$$P_{cond} = \kappa(P_{tot} - P_{rad}) \times e^{-d/\lambda_{pwr}} \qquad (6)$$

where $P_{tot}$ is the total (ohmic and auxiliary) plasma heating power, and $P_{rad}$ is the power radiated in the plasma core, $d$ is the distance between plasma and outer poloidal limiters in the equatorial plane and $\lambda_{pwr}$ is the power decay length. The proportionality factor, κ, and $\lambda_{pwr}$ (~3 cm) were evaluated from heat-flux measurements on the same PFCs in a series of pulses in which antennas A and B were not used and where $d$ was scanned. The fraction of ICRF power dissipated through heating of neighbouring PFCs ranges between 10% and 20 % for +/-π/2 strap phasing and between 2% and 10 % for dipole phasing. This is coherent with the higher ICRF heating efficiency observed when using dipole phasing [22]; however it should be noted that the launched spectrum of the fast wave is different when using dipole (symmetric spectrum centred at $|k_{//,0}|$=6.5m$^{-1}$) or +/-π/2 phasing (asymmetric spectrum centred at $|k_{//,0}|$=3.5 m$^{-1}$), the dipole case having a better single pass absorption in the plasma centre [23]. No clear dependence of $\eta_{ICRF}$ with $P_{ICRF}$ or with plasma-outer limiters distance was observed.

## 5   Conclusions

Simple thermal models of Be tiles, validated with ANSYS, have enabled reliable evaluation of heat-fluxes and characterisation of local heat-loads in the vicinity of energized ICRF antennas. The highest ICRF heat-fluxes observed on JET with the ILW (~4.5 MW/m$^2$ normal flux on PFCs) were obtained when launching 2 MW per antenna with -π/2 phasing and a plasma poloidal-limiter distance of 4 cm. In practice, ICRF operation is not limited by this phenomenon on JET because typical ICRF pulse duration is < 10 s, and because dipole phasing is usually used. However, these heat-flux values deserve further attention, in particular when compared to the engineering design targets (5 MW/m$^2$) for the ITER ICRF antenna Faraday screen and neighbouring blanket modules. Work is on-going to understand the dependence of local RF heat deposition on antenna feeding conditions (number of straps used, voltage balance between the straps and phasing). In particular IR thermography measurements will be compared to RF sheath rectification models taking into account the $E_{//}$ RF electric field distribution at the antenna aperture calculated with the TOPICA code [17] and the field line incidence on the PFC's surface. Finally, the dependencies of the heat-load intensity with antenna voltage indicate that local modifications of the plasma in front of JET ICRF antennas are taking place. Several mechanisms (and sometime with competitive effects) might be involved in this process [24] including RF driven E×B flows, ponderomotive forces, and local ionization of neutrals, which could affect in a complicated manner the parametric dependence of ICRF local heat-loads with system parameters.

**Acknowledgements**

This work, supported by the European Communities under the contract of Association between EURATOM and CCFE, was carried out within the framework of the European Fusion Development Agreement. The views and opinions expressed herein do not necessarily reflect those of the European Commission. This work was also part-funded by the RCUK Energy Programme under grant EP/I501045.

**Figure captions**

Figure 1: Top view of JET, showing the JET ICRF antennas (A, B, C, D, and ILA). The IR camera view is indicated in grey. The poloidal limiters and antenna septa seen by the camera are indicated with arrows. The line of sight of the edge reflectometer (close to the plasma midplane) is also shown.

Figure 2: Wide angle IR camera calibration curve. Calibration data using thermocouple (Tc) measurements from the inner poloidal limiter (crosses) and the outer poloidal limiter (squares) are shown. The dotted lines are the calibration curve when using the inner poloidal Tcs only (top dotted line) or the outer Poloidal Tcs only (bottom dotted lines).

Figure 3: ICRF antenna septum tile thermal modelling. (a) ANSYS model of the tile. The surface temperature response at the centre of the heated area was computed when applying a $2MW/m^2/5s$ heat-flux on the tile surface, the heating patch width is 3 cm (A) or 2.5 cm (B). (b) Surface temperature response as modelled with ANSYS and a simplified 1D model.

Figure 4: (a) Evaluation of the heat-flux on antenna A septum during pulse 81719 with ICRF. The black line is the surface temperature measured by the IR camera. The grey line is the heat-flux evaluated on the surface. (b) Maximum voltage in the transmission line feeding strap 1 (grey line) and strap 2 (black line) of antenna A. (c) Phase difference between forward voltages feeding strap 2 and strap1.

Figure 5: IR view of A and B antennas during pulse 81719. In this pulse only straps 1,2 are energized. Hot-spots can be observed on the antenna septa, and on the poloidal limiters adjacent to the antennas. The area analysed in Figure 4 is encircled.

Figure 6: Maximum heat-flux measured around antenna A plotted vs $n_{e,lim} \times V^2$.

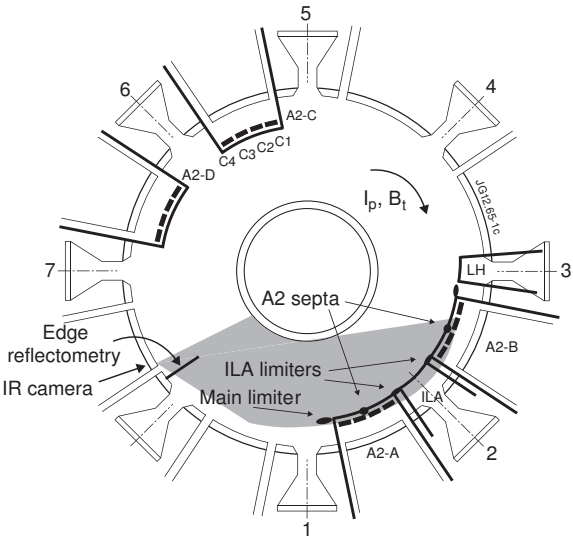

**Figure 1.**

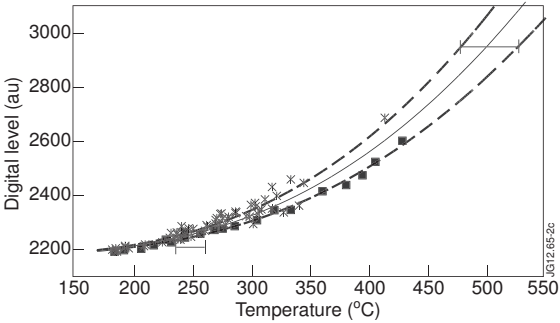

**Figure 2.**

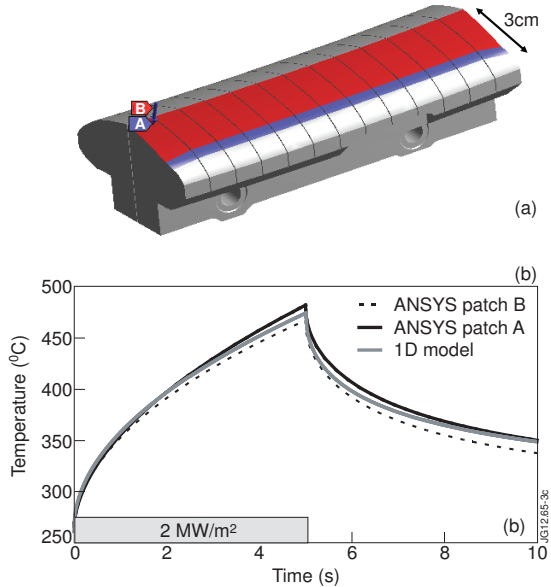

**Figure 3.**

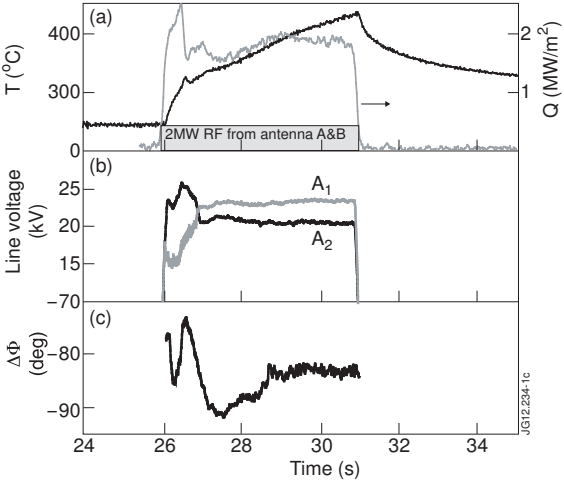

**Figure 4.**

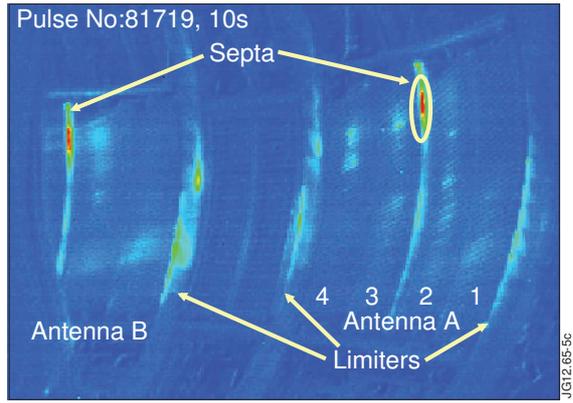

**Figure 5.**

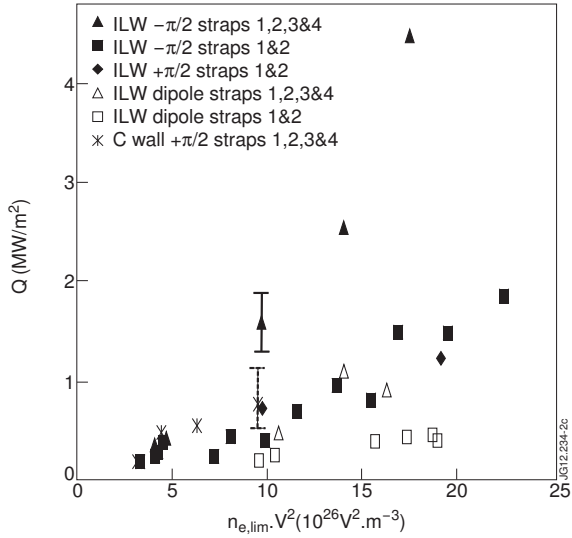

**Figure 6.**